\documentclass[prd, aps,showpacs,prl,twocolumn,noeprint,longbibliography]{revtex4-2}

\usepackage{amsmath}
\usepackage{amssymb}
\usepackage{graphicx}
\usepackage{epsfig}
\usepackage{dcolumn}
\usepackage{textcomp}
\usepackage{bm}
\usepackage[normalem]{ulem} 
\usepackage{color} 
\usepackage{subfigure}
\usepackage{pstricks}
\usepackage{float}
\usepackage{soul}
\usepackage{booktabs}
\usepackage{url}
\usepackage{hyperref}
\usepackage{natbib}
\usepackage[addedmarkup=uwave, defaultcolor=magenta]{changes}

\def\etal{\textit{et~al.\ }}

\newcommand{\ket}[1]{|#1\rangle}
\newcommand{\bra}[1]{\langle #1|}

\DeclareUnicodeCharacter{2009}{\,}

\newcommand{\stkout}[1]{\ifmmode\text{\sout{\ensuremath{#1}}}\else\sout{#1}\fi}
\setdeletedmarkup{\stkout{#1}}

\begin{document}

\title{The influence of pinholes and weak-points in aluminium-oxide Josephson junctions} 

\author{K. Bayros, M. J. Cyster, J. S. Smith}
\author{J. H. Cole}
\email{jared.cole@rmit.edu.au}
\affiliation{%
Chemical and Quantum Physics, School of Science, RMIT University, Melbourne, Victoria 3001, Australia.
}



\date{\today}
             
\begin{abstract}
Josephson junctions are the key components used in superconducting qubits for quantum computing. The advancement of quantum computing is limited by a lack of stability and reproducibility of qubits which ultimately originates in the amorphous tunnel barrier of the Josephson junctions and other material imperfections. Pinholes in the junction have been suggested as one of the possible contributors to these instabilities, but evidence of their existence and the effect they might have on transport is unclear. We use molecular dynamics to create three-dimensional atomistic models to describe Al$-$AlO$_x-$Al tunnel junctions, showing that pinholes form when oxidation of the barrier is incomplete. Following this we use the atomistic model and simulate the electronic transport properties for tunnel junctions with different barrier thicknesses using the non-equilibrium Green's function formalism. We observe that pinholes may contribute to excess quasiparticle current flow in Al$-$AlO$_x-$Al tunnel junctions with thinner barriers, and in thicker barriers we observe weak-points which facilitate leakage currents even when the oxide is continuous. We find that the disordered nature of the amorphous barrier results in significant variations in the transport properties. Additionally, we determine the current-phase relationship for our atomistic structures, confirming that devices with pinholes and weak-points cause a deviation from the ideal sinusoidal Josephson relationship.
\end{abstract}

\maketitle

\section{Introduction}
Josephson junctions are one of the key components used in superconducting qubits for quantum computers \cite{Nielsen2010, Zagoskin2011, Wendin2007}. Whilst there have been significant developments in the fabrication processes of Josephson junctions which now enable us to have working examples of small scale quantum computing \cite{Chen2021, Kim2023, Kjaergaard2020}, we still see significant problems with device stability and reproducibility. In particular, large variabilities have been found in the coherence times with time \cite{Carroll2022, Muller2015}, and across different qubits on a quantum computing chip \cite{Kreikebaum2020, Osman2021, Verjauw2022}. Variations have also been found in the critical currents between Josephson junctions fabricated in the same way \cite{VanHarlingen2004}. These variations in the qubit parameters means that significant tuning is required to ensure qubits across the chip function similarly. Although qubits can be tuned \cite{Werninghaus2021, Wittler2021}, this typically requires more gates to be patterned onto quantum computing chips, as well as recalibration during the computation \cite{Kjaergaard2020, Hazard2023}. This significantly limits the reliability of quantum computing chips that are fabricated today. It is not yet clear what causes stability and reproducibility problems in these devices, however a large amount of research discusses the importance of the tunnel barrier in this context \cite{Muller2019}. Often comprised of an amorphous metal-oxide material, the tunnel barrier is inherently disordered, inviting the possibility of structures that could act as two-level systems that couple to the qubit current causing instabilities in its parameters. A particular type of imperfection in the tunnel barrier that has been proposed are microscopic metallic links, coined ``pinholes". These pinholes may increase the current that couples to two-level systems in the barrier, exacerbating instabilities \cite{Constantin2007}. 

In the literature there is currently conflicting evidence on the role and prevalence of pinholes. Zhou \etal \cite{Zhou2007} finds that pinholes may exist in particular circumstances, such as when the oxide barrier is too thin, whereas Greibe \etal \cite{Greibe2011} develops a model which rules out pinholes as the source of excess current. On the other hand, Tolpygo \etal \cite{Tolpygo2008} shows that pinholes may not exist in a newly fabricated device, however they may form through degradation of the oxide when the junction is used in a circuit. Recent work has studied the effect of Josephson harmonics on the current-phase relationship (CPR), highlighting that the disordered nature of the oxide barrier could contribute to effects in the CPR, specifically if there are  high transmission channels present \cite{Willsch2024}.

This paper uses computational techniques to study how pinholes form and aims to investigate what effects they might have on transport properties in Josephson junction devices that are fabricated today. Commonly, Josephson junctions are fabricated as Al$-$AlO$_x-$Al tunnel junctions \cite{Zeng2016,Schulze1998,Morohashi1985}. This study investigates pinhole formation during AlO$_x$ growth to discuss under which conditions they might be stable, and what effect stable pinholes might have on electronic transport through the device. We begin by discussing the stability of pinholes following Zhou \etal \cite{Zhou2007} which models a cylindrical pinhole in an otherwise uniform oxide barrier. Using the interfacial surface energies between materials in the system the stability of pinholes can be determined. Using a similar method we determine the stability of cylindrical pinholes with increasing radii for different barrier thicknesses. Using the same geometries we then calculate the transport properties for each of these devices.

To then investigate the formation and closing of pinholes more realistically, we use our previously developed molecular dynamics approach \cite{Cyster2021} to grow Al$-$AlO$_x-$Al tunnel junctions with different amounts of oxide in the barrier.

Using a tight-binding description of the system we employ a non-equilibrium Green's function (NEGF) approach to investigate normal (single-electron) transport in the structures studied herein. NEGF is a numerical approach which has been widely used for calculating transport properties in nanoscale devices \cite{Datta2005}, and has also been used explicitly for the case of Al$-$AlO$_x-$Al tunnel junctions \cite{Cyster2020}. In this work, we determine the transmission, resistance-area, and the current density through full three-dimensional models of Al$-$AlO$_x-$Al tunnel junctions with different amounts of oxide. Understanding variations in the normal transport based on structural irregularities can suggest causes for variation in the superconducting properties.

To understand how the normal transport translates to the superconducting (correlated-electron) transport, the Ambegaokaar-Baratoff (AB) relation is commonly used \cite{Tinkham2004, Heikkila2013}. It specifically relates normal-state resistance to the superconducting critical current. The validity of this relationship is unclear when there are pinholes present in the oxide barrier. As an alternative the model in this work is used to calculate the transmission eigenvalues for each of our structures, which in turn allows us to determine the contributions from different transmission channels and the effects they have on the CPR.

From a toy model we determine that pinholes may be present in thinner barriers and a source of single-electron current. We then use the analysis from the toy model to understand the transport properties of a three-dimensional atomistic model describing Al$-$AlO$_x-$Al tunnel junctions derived from molecular dynamics simulations. Our results show there is a significant amount of variability in the transport properties. These variabilities are still present for devices without pinholes, indicating that the variation is a product of the inherently disordered AlO$_x$ barrier. We also find that the amorphous structure of the oxide results in a non-uniform electric potential, which makes the metal-insulator transition of the barrier less clear compared to analysis from our toy model. Specifically, some results highlight that thick barriers with no pinholes may remain conductive. These devices do not have clear pinholes in the barrier, but have weak-points which can facilitate single-electron current flow. This could indicate a situation in which a seemingly complete barrier could result in a dysfunctional device.

\section{Toy model}
\begin{figure}[htb]
	\centering
	\includegraphics[scale=1]{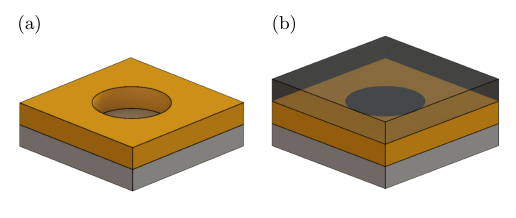}
	\caption{A schematic representation of a simple cylindrical pinole in an Al$-$AlO$_x-$Al tunnel junction. (a) Aluminium bottom contact (grey) and an aluminium-oxide layer (orange) with a pinhole. (b) A complete tunnel junction showing the aluminium-oxide layer with a pinhole underneath an aluminium top contact (transparent grey).}
	\label{fig:pinhole}
\end{figure}

Given the complex structure of disordered AlO$_x$, we first develop an idealised model to understand how pinholes of a simple structure remain stable in the barrier and also how systematic changes in pinhole size may affect the transport properties of tunnel junctions. For this section we use a geometric model with pinholes described as cylinders in an otherwise uniform oxide layer in Al$-$AlO$_x-$Al tunnel junctions. Fig.~\ref{fig:pinhole} schematically depicts the geometry of our model.
 
\subsection{Stability of pinholes}
To study under which conditions pinholes are stable in the oxide barrier, we follow methods outlined in Zhou \etal \cite{Zhou2007} to determine the interfacial surface energies between different materials in our system. Previous theoretical studies use molecular dynamics methods to calculate the interfacial surface energies for different Al and AlO$_x$ interfaces \cite{Zhou2005}, finding $\gamma_\mathrm{Al} = 0.057$ eV/{\AA}$^2$, $\gamma_\mathrm{AlO} = 0.341$ eV/{\AA}$^2$, and $\gamma_\mathrm{Al/AlO} = 0.186$ eV/{\AA}$^2$ which we use in subsequent calculations. The surface energy,  $\gamma$, is between the materials specified in the subscript, and where only one material is given, the interface is vacuum.

Eq.~\ref{eq:formation_energy} describes the hole formation energy for the geometry shown in Fig.~\ref{fig:pinhole}~(a), with a pinhole in the oxide on an Al bottom contact and vacuum above the oxide (before the top contact is deposited).
\begin{multline}
\label{eq:formation_energy}
\Delta E = \pi r^2 (\gamma_\mathrm{Al} - \gamma_\mathrm{AlO} - \gamma_\mathrm{Al/AlO}) + 2 \pi r h (\gamma_\mathrm{AlO})
\end{multline}

\begin{figure}[htb]
	\centering
	\includegraphics[scale=1]{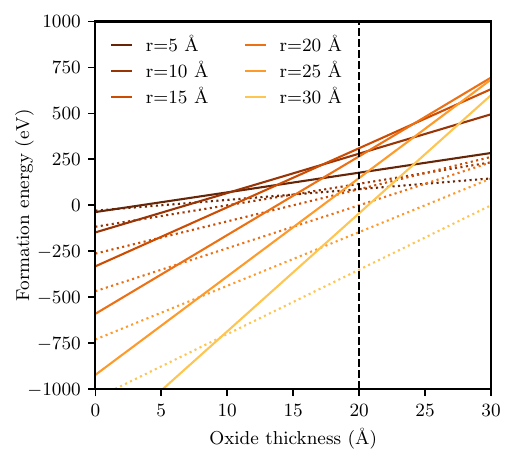}
	\caption{The energy required to form cylindrical pinholes with different radii in AlO$_\text{x}$ barriers of different thicknesses. Solid lines indicate a calculation with vacuum above the oxide layer, dashed lines show a calculation with an aluminium top contact. Black dashed line indicates a typical oxide thickness off 20 {\AA} used in Al$-$AlO$_x-$Al junctions.}
	\label{fig:formation_energy}
\end{figure}

A similar equation can be derived for the case with an Al top contact deposited. Using Eq.~\ref{eq:formation_energy}, we can determine the energy required to form holes of different radii, $r$, in the oxide barrier at different thicknesses, $h$. Fig.~\ref{fig:formation_energy} shows this relationship for the two geometries shown in Fig.~\ref{fig:pinhole}.

A negative hole formation energy implies that for a uniform oxide layer arrangement it is more energetically favourable to have pinholes and they may form spontaneously. On the other hand, a positive hole formation energy indicates that energy is required to form pinholes in a uniform oxide layer. The results in Fig.~\ref{fig:formation_energy} show that for thinner oxides, it is more energetically favourable for larger pinholes to form. However as the oxide is grown thicker there is a cross-over point where it becomes more energetically favourable to form smaller pinholes. This is consistent with the oxide beginning to cover more of the underlying substrate, resulting in the formation of smaller holes. Although this is the case, it is more common that the formation energy becomes positive before we see this cross over, indicating it is not favourable for pinholes of any size to form.

For the case of a complete junction as depicted in Fig.~\ref{fig:pinhole}~(b) where aluminium is deposited as a top contact, we find that the formation energy remains negative for larger oxide thicknesses. This implies that thicker oxide barriers are required to prevent the spontaneous formation of pinholes compared to the case when the oxide is exposed to vacuum. Consequently, this means that the deposition of aluminium on top of existing pinholes may increase their stability.

The results in Fig.~\ref{fig:formation_energy} are consistent with previous findings in work on dewetting of thin-films where it has been found that the number of holes formed scales inversely with the film thickness \cite{Presland1972}. This highlights the importance of considering the stability of pinholes when studying formation of Al$-$AlO$_x-$Al tunnel junctions, specifically ones with thinner barriers.

\subsection{Normal transport in toy model}
\label{sec:toy_model}

Using a model with the same geometry as depicted in Fig.~\ref{fig:pinhole}~(b), we develop a three-dimensional tight-binding model. We do this here to compare to later sections where this method is used with a full atomistic description of the device (see Sec.~\ref{sec:md_transport}). For all our calculations we take $z$ to be the direction of transport, so that the $x-y$ plane defines the cross-sectional area of the device. We define a potential in three-dimensions with dimensions in $x$ and $y$ of 24 {\AA} $\times$ 24 {\AA}, and a barrier height and length of 1.3 eV and 30 {\AA} respectively. When Al$-$AlO$_x-$Al junctions are fabricated they can range from $0.03 - 1$~$\mu$m$^2$ in cross sectional area \cite{Bilmes2022, Mamin2021}, which is significantly larger than our simulated device area. To account for this we apply periodic boundary conditions in $x$ and $y$, whereas $z$ has open boundary conditions to simulate transport \cite{Cyster2020}.

To calculate the normal transport through the junction we employ a non-equilibrium Green's function (NEGF) approach which is a widely used method for solving the time-independent Schr\"{o}dinger equation (TISE) numerically for a system with open boundary conditions \cite{Datta2005,Datta1997}. This allows us to simulate the effects of attaching a source and drain contact to the barrier and subsequently  calculate properties such as the transmission and normal state resistance through the device.

The retarded Green's function is given by
\begin{equation}
G^r(E) = [(E+i\eta)I-H-\Sigma_\textrm{S}-\Sigma_\textrm{D}]^{-1}
\end{equation}

where $I$ is the identity matrix, and $i\eta$ is a positive imaginary infinitesimal number.

The Hamiltonian, $H = T + U$, is defined with a kinetic, $T$, and potential, $U$, energy operator. We describe our system using a second-order finite difference representation of the kinetic energy operator,
\begin{equation}
T = \sum_{i}^N \varepsilon \ket{i}\bra{i} - \sum_{<i,j>}^{N} t_k \ket{i}\bra{j} \:,
\end{equation}

where $\varepsilon = 2t_x + 2t_y + 2t_z$, and $k \in \{x,y,z\}$. The hopping parameter, $t_k$, is given by $t_k = \frac{\hbar^2}{2m^*a_k^2}$ where $m^*$ is the effective mass and $a_k$ is the finite grid spacing in a given direction. We assume the effective electron mass $m^*=9.1\times10^{-31}$~kg as the structures considered are mostly pure aluminium and we use $a_k=1/3$~{\AA} throughout as this has previously shown good convergence for these structures \cite{Cyster2020}. In this section the electrostatic potential energy term, $U$, is applied as a rectangular barrier with a cylindrical hole in it. In Sec. \ref{sec:atomistic_model}, $U$ is determined from the charges and positions of the atoms. The self energies for the source and drain, $\Sigma_\textrm{S/D}$, are calculated recursively \cite{Ozaki2010a} and these terms account for the effect of attaching semi-infinite leads.

The normal transmission is then calculated from \cite{Datta2005}
\begin{equation}
\label{eq:transmission}
T(E) = \textrm{Tr}(\Gamma_\textrm{S} G^r \Gamma_\textrm{D} G^a)
\end{equation}

where the broadening matrices are given by ${\Gamma_{\textrm{S,D}} = i(\Sigma_{\textrm{S,D}} - \Sigma_{\textrm{S,D}}^\dagger)}$, and the advanced Green's function is $G^a = G^{r \dagger}$.

The Landauer-B{\"u}ttiker formula allows us to use $T(E)$ to calculate the current in a device:
\begin{equation}
\label{eq:landauer}
I = \frac{2e^2}{h} \int_{-\infty}^{\infty} T(E) [ f_\textrm{S}(E) - f_\textrm{D}(E)] \textrm{d}E \:
\end{equation}

where the Fermi-Dirac distributions for the source and drain contacts are given by ${f_{\textrm{S,D}} = 1/[\textrm{exp}((E - E_\textrm{F} - eV_{\textrm{S,D}}/2)/k_\textrm{B} T) + 1]}$, $e$ is the electron charge, $h$ is Planck's constant, $E_\textrm{F}$ is the Fermi energy, $k_\textrm{B}$ is the Boltzmann constant, and $T$ is the temperature.

The normal state resistance at a given energy is determined by taking the zero-temperature, zero-bias limit of Eq.~\ref{eq:landauer} and is given by
\begin{equation}
\label{eq:normal_resistance}
R_\textrm{N} = \frac{h}{2e^2} \frac{1}{T(E=E_\textrm{F})} \:.
\end{equation}

In general, modelling superconductivity with a NEGF method is more difficult than for normal transport \cite{Sriram2019,Datta1997}. However, we are able to perform an analysis of the transmission channels and thereby estimate the superconducting response in Sec.~\ref{sec:toy_sc}.

\begin{figure}[htb]
	\centering
	\includegraphics[scale=1]{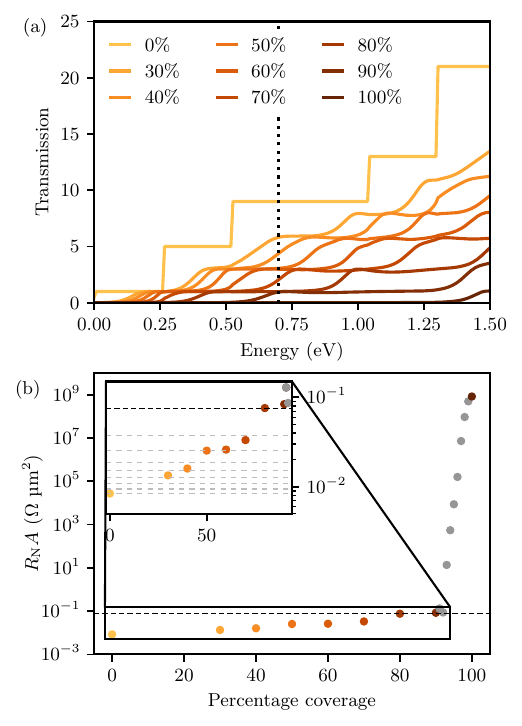}
	\caption{(a) Transmission through cylindrical pinholes in a 3D rectangular barrier which leave different percentages of the underlying substrate covered, 0\%, and 30\%-100\% in intervals of 10\%. The vertical dashed line indicates $E_\textrm{F} = 0.7$ eV. (b) The corresponding resistance-area products calculated at $E_\textrm{F}$ for each of these junctions with extra points shown between 90\%-100\% to show the metal to insulator transition. The inset shows a zoomed in area indicated by the rectangle, with dashed lines corresponding to resistance quanta.}
	\label{fig:toy_transport}
\end{figure}

Fig.~\ref{fig:toy_transport} shows the transmission and resistance-area product through junctions with pinholes of different radii. To ensure consistency and clarity in later sections, we henceforth define the presence of pinholes using the percentage of the underlying substrate that is covered. In this section we model pinholes as cylinders, so our substrate coverage begins at 30\% as this allows a full circle to fit within the square cross-sectional area of the device that is simulated. 

The $x$ and $y$ dimensions of the potential barrier used for results in Fig.~\ref{fig:toy_transport} are chosen to compare to an example barrier from molecular dynamics calculations which are performed in Sec.~\ref{sec:atomistic_model}. Similarly we choose the barrier height and thickness to correspond to the thickest barriers in Sec.~\ref{sec:atomistic_model}. We define the barrier region as the distance from the first oxygen atom to the last oxygen atom in the structure, and then find the average height in the middle 50\% of the barrier to be 1.3 eV. Integrating the electric potential in the barrier region we find the barrier area and consequently determine the effective width of the barrier as 30 {\AA}.

In Fig.~\ref{fig:toy_transport}~(a) we see quantised transmission steps which show the modes of conduction available. As more of the underlying substrate is covered and the barrier becomes more complete, we see smoothing of these quantised steps. At the extremes, the 0\% coverage data shows ballistic conduction, and the 100\% indicates transmission through a uniform rectangular barrier. We also see an increase in the transmission at all energies as the percentage coverage decreases, indicative of a more conductive device as is expected when pinholes are present in barrier.

In order to calculate the resistance using Eq.~\ref{eq:normal_resistance} we require the Fermi energy, $E_\textrm{F}$, of the leads. In order to determine a suitable $E_\textrm{F}$ we use measurements from prior experiments where the resistance-area product was found to be $R_\textrm{N}A = 600$ $\Omega$~\textmu m$^2$ \cite{Aref2014}. Using this value as a standard, we calculate the corresponding transmission value $T = 4 \times 10^{-5}$ using Eq.~\ref{eq:normal_resistance} and the quoted device cross-sectional area. Linearly interpolating between points in the transmission function of our 100\% coverage data set, we determine the corresponding $E_\textrm{F} \approx 0.7$ eV which we use for all subsequent calculations.

Fig.~\ref{fig:toy_transport}~(b) shows $R_\textrm{N}A$ calculations using $E_\textrm{F} = 0.7$ eV, marked with a dashed line with the transmission results in Fig.~\ref{fig:toy_transport}~(a). Between 90-100\% coverage we have included extra points in intervals of 1\% to highlight the sharp increase in resistive behaviour as we approach 100\% coverage. This jump corresponds to the transition from metal conduction to insulating barrier. To demonstrate this clearly, the inset in Fig.~\ref{fig:toy_transport}~(b) shows horizontal dashed lines at multiples of one resistance quantum. The darkest dashed line corresponds to $T = 1$, or a resistance value of $R_\textrm{N} A = 0.07$ $\Omega$~\textmu m$^2$ which is also marked on the main figure for clarity. Beyond this there are no modes to facilitate ballistic conduction and the device becomes insulating, i.e.~all conduction occurs via barrier tunnelling.


\subsection{Superconducting transport in toy model}
\label{sec:toy_sc}
The supercurrent-phase relationship \cite{Heikkila2013,Haberkorn1978,Beenakker1991} in a Josephson junction is given by
\begin{equation}
\label{eq:supercurrent}
I_\textrm{S}(\phi) = \sum\limits_p \frac{e \Delta^2}{2 \hbar} \frac{\tau_p \sin(\phi)}{E_{+}^{\textrm{ABS}}} \tanh \left[ \frac{E_{+}^{\textrm{ABS}}(\phi)}{2k_\textrm{B} T} \right]
\end{equation}
where
\begin{equation}
E_{\pm}^{\textrm{ABS}} = \pm \Delta\sqrt{1-\tau_p \sin^2(\phi/2)} \:.
\end{equation}

Here, $\tau_p$ is the transmission probability for a transmission eigenmode $p$, $\Delta$ is the superconducting order parameter, and $\hbar$ is the reduced Planck constant. $E_{\pm}^{\textrm{ABS}}$ is the energy of the Andreev bound states which mediate Cooper pair transfer between superconducting regions, leading to the Josephson effect. Eq.~\ref{eq:supercurrent} is valid for short Josephson junctions, i.e.~when the device length is much less than the superconducting coherence length \cite{Tinkham2004}, $L \ll \xi$. Aluminium has a long coherence length of $\xi = 1600$ nm \cite{Kittel1955} which makes it widely suitable for Josephson junction applications as the device lengths are typically much shorter.

If $\tau_p$ is low for all transmission eigenmodes ($\tau_p \ll 1$) the supercurrent-phase relationship in Eq.~\ref{eq:supercurrent} can be written as the Ambegaokaar-Baratoff (AB) relation,
\begin{equation}
\label{eq:ab_relation}
I_\textrm{S} R_\textrm{N} = \frac{\pi \Delta \sin(\phi)}{2e} \text{tanh} \left( \frac{\Delta}{2k_\textrm{B} T} \right)
\end{equation}
where $R_\textrm{N} = h/[2e^2 \sum_{p} \tau_p$]. This equation relates the transport properties of the device in its normal state to properties of the device in its superconducting state. In a similar manner, in the fully transparent limit where all $\tau_p = 1$, Eq.~\ref{eq:supercurrent} can be written as the Kulik Omel'yanchuk (KO) relation \cite{Golubov2004} which is given by

\begin{equation}
\label{eq:ko_cpr}
I_\textrm{S}R_\textrm{N} = \frac{\pi\Delta \sin(\phi/2)}{e} \tanh \left( \frac{\Delta \cos{\phi/2}}{2 k_\textrm{B} T} \right) \:.
\end{equation}

Taking the zero-temperature limit of Eq.~\ref{eq:ko_cpr},  $T \to 0$,  we find $\tanh \left(  \Delta \cos \phi/2/[2k_\textrm{B}T] \right) \to \textrm{sign}(\cos \phi/2)$. Typically the relations in Eq.~\ref{eq:ab_relation} and \ref{eq:ko_cpr} are used for devices with a complete oxide barrier \cite{Nugroho2013}, however in the case of devices with pinholes and weak-points there are channels with different transmission probabilities. This results in a combination of channels which display characteristic AB and KO behaviour, as well as behaviour in between. Summing these channels then causes a deviation from the ideal sinusoidal Josephson relationship.

To determine the transmission probability eigenvalues \cite{Paulsson2007, Burkle2012} in our model, we find the largest thirty eigenvalues of the transmission matrix, $\Gamma_\textrm{S} G^r \Gamma_\textrm{D} G^a$, which is also used in Eq.~\ref{eq:transmission} to determine the normal transmission.

\begin{figure}[H]
	\centering
	\includegraphics[scale=1]{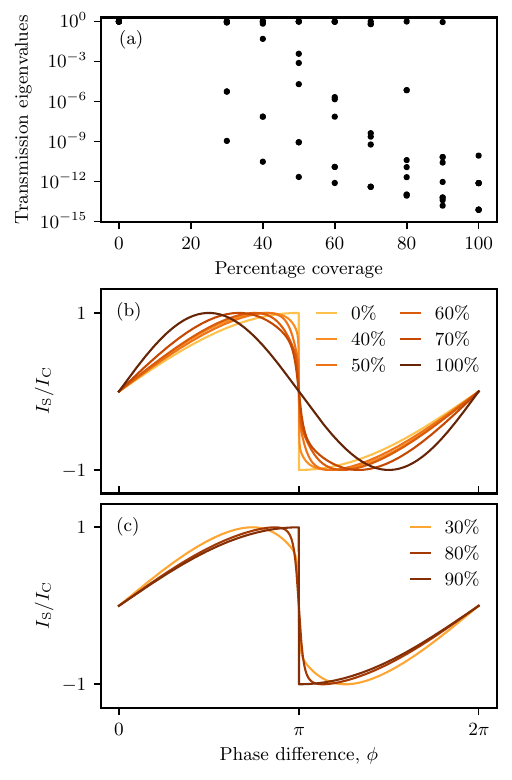}
	\caption{(a) Eigenvalues of the transmission matrix for the toy model structures 0\%, and 30\%-100\% coverage in 10\% intervals. (b) CPRs for structures indicated in the legend. (c) CPRs which correspond to the low and high coverage regimes, for structures indicated in the legend.}
	\label{fig:toy_cpr}
\end{figure}

We take the low-temperature limit of Eq.~\ref{eq:supercurrent} to calculate the supercurrent for each transmission eigenvalue. We use a low-temperature superconducting gap value of $\Delta=200$ \textmu eV as measured for thin films in prior work \cite{Cedergren2015}. Fig.~\ref{fig:toy_cpr} shows the sum of the supercurrent from each of these channels, giving the overall CPR.

\begin{figure*}[htb]
	\centering
	\includegraphics[scale=0.55]{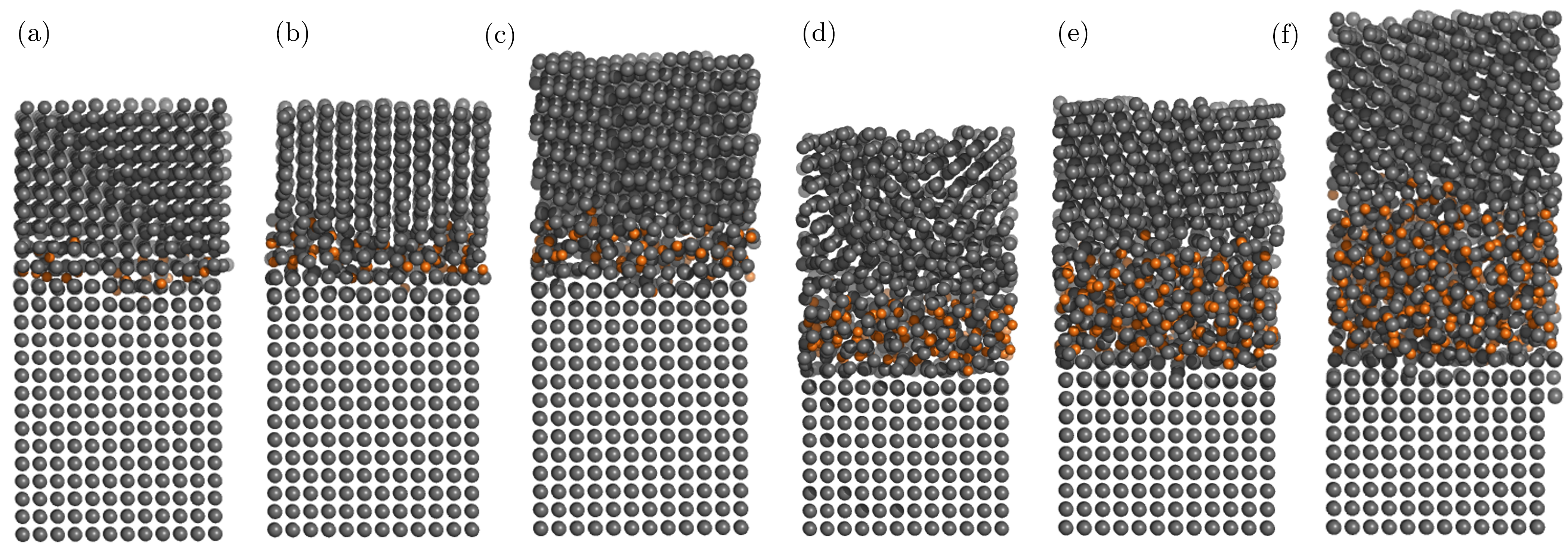}
	\caption{Molecular dynamics simulation for a `(100)1' junction with increasing amounts of oxide in the barrier. Aluminium atoms are represented in grey, and oxygen atoms in orange. Structures are presented in order of number of oxygen atoms in the barrier, (a-f) \# O = 28, 56, 83, 143, 240, 346. See Table~\ref{tab:thickness} for more details on oxide thicknesses.}
	\label{fig:coverage}
\end{figure*} 

Fig.~\ref{fig:toy_cpr}~(a) shows the eigenvalues for the transmission matrix for each toy model structure.  Notably, we find that the the eigenvalues in the 0\% coverage structure are all approximately equal to 1. With increasing coverage some eigenvalues begin to drop closer to zero, indicating channels with lower transmission probabilities are present. For the structure with 100\% coverage, there are no channels with transmission probabilities close to 1. In Fig.~\ref{fig:toy_cpr}~(b) we see that for a complete barrier (100\% coverage) we obtain the expected sinusoidal Josephson relationship for ideal Josephson junctions described by the AB relation (Eq.~\ref{eq:ab_relation}). In the other extreme for ballistic conduction, or in the fully transparent limit, we obtain a sawtooth relationship which is given by the KO relation (Eq.~\ref{eq:ko_cpr}). Aside from these extremes, for percentage coverages between 40-70\% the current-phase relationship follows the expected trend from sawtooth to sinusoidal for increasing surface coverages.

We find exceptions to this in the very low-coverage and very high-coverage cases, given by the 30\%, 80\%, and 90\% coverage data sets shown in Fig.~\ref{fig:toy_cpr}~(c). These cases correspond to two different regimes. In the very low-coverage regime indicated by the 30\% coverage data set, the diameter of the cylindrical pinhole is approximately equal to the width of the substrate in our simulation. Due to the periodic boundary conditions applied in our model, this creates a pattern which is similar to diamonds of oxide on a metallic substrate, rather than a hole in a uniform oxide layer. This pattern leads to a shape for the CPR that does not fall between the results for 0\% and 40\% coverage. In the high-coverage cases we approach a limit where we see a single transmission channel that dominates the conduction. In Fig.~\ref{fig:toy_cpr}~(a) there are multiple eigenmodes with a transmission probability close to 1, however for 80\% and 90\% there is only one eigenvalue. This is resemblant of the Josephson effect for a narrow constriction, or a superconducting quantum point contact \cite{Furusaki1992}.

We use the insights from this toy model study to inform our interpretation of results in later sections.

\section{Atomistic model}
\label{sec:atomistic_model}

\subsection{Molecular dynamics}
\label{sec:md}

In the preceding sections pinholes have been described as cylindrical holes in the oxide barrier, however in practice pinholes are likely to have irregular shapes. To represent pinhole formation in Al$-$AlO$_x-$Al tunnel junctions more accurately we use molecular dynamics to grow devices atomistically, allowing us to resolve microscopic structures within the barrier. In order to simulate this growth, we follow a similar method to that used by Cyster \etal \cite{Cyster2021}, and in this work we use the Large-scale Atomic/Molecular Massively Parallel Simulator (LAMMPS) \cite{Plimpton1995, LAMMPS} to perform our simulations with a ReaxFF force field to describe atom to atom interactions \cite{VanDuin2001, Aktulga2012}. The methods in Cyster \etal are designed to replicate the Dolan double-angle evaporation process \cite{Dolan2008}, and model the low-pressure oxidation and aluminium evaporation processes that are often used when fabricating Al$-$AlO$_x-$Al tunnel junctions. 

Beginning with a 24 {\AA} $\times$ 24 {\AA} substrate cell of crystalline aluminium, we introduce a vacuum above the substrate and grow an AlO$_x$ layer by introducing oxygen atom-by-atom with atom trajectories and velocities sampled from a Boltzmann distribution. We create a set of twelve structures with increasing numbers of oxygen atoms in the AlO$_x$ layer, some of which are shown in Fig.~\ref{fig:coverage}. We find that the layers with less oxide tend to contain pinholes which close over as more oxide is grown (see Sec.~\ref{sec:characterising}). 

We find that this simulation method results in self-limiting AlO$_x$ growth when the LAMMPS/ReaxFF forcefield is used (in contrast to our previous work with GULP/Streitz and Mintmire (S-M) \cite{Cyster2020} which did not self-limit). However the resulting oxide grown with LAMMPS/ReaxFF is thinner (approximately 1 {\AA}) \cite{Aref2014,Fritz2019} than is typically observed experimentally. To study thicker oxides we replicate the methods used to grow \textgreater10~nm thick oxides experimentally. Namely, after the oxide self-limits, an atomically thin layer of aluminium is deposited atom-by-atom and is oxidised following the same process as before. This is repeated until the oxide reaches the desired thickness. 

After the oxidation an aluminium contact is deposited on top in a similar manner, introducing aluminium atoms one at a time until the aluminium contact layer is of a similar thickness to the initial aluminium substrate. In this case, the distribution of velocities is chosen to mimic the metal evaporation, see Cyster \etal \cite{Cyster2021} for details. The structure is then optimised using the LAMMPS energy minimisation to find the lowest energy configuration of the atoms that constitute the junction. We calculate the charges of the atoms in the system and then the Ewald summation method can ultimately be used to determine the electric potential of the system with respect to atom position \cite{Ewald1921}.

This method is used with an Al (100) and Al (111) substrate for two complete simulations of each, with varying thicknesses of oxide giving 48 structures in total as shown in Table~\ref{tab:thickness}.

\subsection{Characterising the junction formation}
\label{sec:characterising}
\subsubsection{Thickness}

\begin{figure}[htb]
	\centering
	\includegraphics[scale=1]{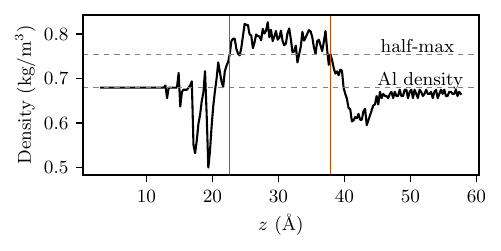}
	\caption{The rolling average of the mass density across the thickest barrier in the `(100)1' data set. The half-max reference line is halfway between the maximum value in the data set and the Al density, where the latter is set using the mass density value  at the beginning of the source contact. The solid orange lines then indicate the FWHM which we take as the barrier thickness.}
	\label{fig:FWHM}
\end{figure}

When simulating oxide growth, we require a measure of thickness to indicate when to stop the simulation. For this, we use the $z$ distance from the first to last oxygen atom in the junction (oxygen extent) as an approximate guide to the barrier thickness. Table~\ref{tab:thickness} shows these barrier thicknesses for each structure. We have also determined the full-width half-maximum (FWHM) thickness of the barrier for each structure. This is calculated using the maximum of the average mass density across the junction shown in Fig.~\ref{fig:FWHM}.

\begin{table*}[t]
\begin{center}
\caption{Summary of parameters used to characterise junction formation in this work for each data set. The number of oxygen atoms in a simulation cell is given by `\# O', and the percentage coverage calculated using the Voronoi cells is `\% coverage'. The thicknesses from the FWHM of the mass density (shown in Fig.~\ref{fig:FWHM}), and the oxygen extent are labelled `FWHM' and `O extent' respectively.}
\label{tab:thickness}

	\begin{tabular}{ c c c c c c c c c c c c c c c}
		\toprule
		
		 \multicolumn{3}{c}{(100)1} & \hspace{0.7cm} & \multicolumn{3}{c}{(100)2} & \hspace{0.7cm} & \multicolumn{3}{c}{(111)1} & \hspace{0.7cm} & \multicolumn{3}{c}{(111)2} \\
		 
		 \midrule
		 
		 \# O & \multicolumn{2}{c}{\% coverage} & & \# O & \multicolumn{2}{c}{\% coverage} & & \# O & \multicolumn{2}{c}{\% coverage} & & \# O & \multicolumn{2}{c}{\% coverage} \\
		 
		\cmidrule(r{4pt}){1-3} \cmidrule(r{4pt}){5-7} \cmidrule(r{4pt}){9-11} \cmidrule(r{4pt}){13-15}
		
		28 & \multicolumn{2}{c}{27.81} & & 20 & \multicolumn{2}{c}{24.28} & & 26 & \multicolumn{2}{c}{25.38} & & 24 & \multicolumn{2}{c}{27.38} \\
		36 & \multicolumn{2}{c}{36.88} & & 26 & \multicolumn{2}{c}{30.34} & & 33 & \multicolumn{2}{c}{32.74} & & 29 & \multicolumn{2}{c}{32.37} \\
	    56 & \multicolumn{2}{c}{52.77} & & 43 & \multicolumn{2}{c}{46.07} & & 46 & \multicolumn{2}{c}{42.75} & & 32 & \multicolumn{2}{c}{35.76} \\
		69 & \multicolumn{2}{c}{59.24} & & 54 & \multicolumn{2}{c}{52.92} & & 56 & \multicolumn{2}{c}{51.70} & & 44 & \multicolumn{2}{c}{46.38} \\
		83 & \multicolumn{2}{c}{68.48} & & 68 & \multicolumn{2}{c}{67.06} & & 84 & \multicolumn{2}{c}{63.90} & & 58 & \multicolumn{2}{c}{59.08} \\
		139 & \multicolumn{2}{c}{92.12} & & 148 & \multicolumn{2}{c}{91.31} & & 140 & \multicolumn{2}{c}{91.68} & & 120 & \multicolumn{2}{c}{86.69} \\
		
		\cmidrule(r{4pt}){1-3} \cmidrule(r{4pt}){5-7} \cmidrule(r{4pt}){9-11} \cmidrule(r{4pt}){13-15}
		
		 \# O & FWHM & O extent & & \# O & FWHM & O extent & & \# O & FWHM & O extent & & \# O & FWHM & O extent \\
		 
		\cmidrule(r{4pt}){1-3} \cmidrule(r{4pt}){5-7} \cmidrule(r{4pt}){9-11} \cmidrule(r{4pt}){13-15}
		 
		143 & 4.75 & 10.07 & & 150 & 3.00 & 10.25 & & 146 & 4.75 & 10.57 & & 134 & 4.25 & 10.46 \\
		239 & 6.00 & 12.74 & & 158 & 3.25 & 12.44 & & 156 & 4.00 & 12.72 & & 206 & 7.25 & 13.40 \\
		240 & 6.00 & 14.21 & & 243 & 6.75 & 15.50 & & 232 & 7.00 & 14.03 & & 245 & 8.00 & 14.77 \\
		311 & 8.75 & 16.48 & & 250 & 6.00 & 16.81 & & 248 & 9.50 & 16.62 & & 269 & 7.75 & 16.64 \\
		346 & 9.50 & 20.37 & & 255 & 6.00 & 22.89 & & 279 & 10.00 & 19.03 & & 306 & 9.00 & 20.66 \\
		451 & 15.25 & 23.49 & & 383 & 13.50 & 23.49 & & 397 & 15.50 & 24.10 & & 399 & 14.50 & 24.09 \\
		
		\bottomrule
	\end{tabular}
\end{center}
\end{table*}

Due to the lower stoichiometric ratio of Al:O near the Al/AlO$_x$ interface, we expect the oxygen extents in Table~\ref{tab:thickness} to be an overestimation of the thickness that would be measured experimentally. We include the FWHM thickness computed from the mass density as an alternate measure, however this in turn may be an underestimation. Given the ambiguity of defining oxide thickness at the atomic scale, Table~\ref{tab:thickness} gives both values as an upper and lower limit of the thickness for a given structure. The intervals between FWHM values are multiples of 0.25 due to the bin size chosen for the rolling average of the mass density.

\subsubsection{Percentage coverage}
\label{sec:perc_coverage}

\begin{figure}[b]
	\centering
	\includegraphics[scale=1.1]{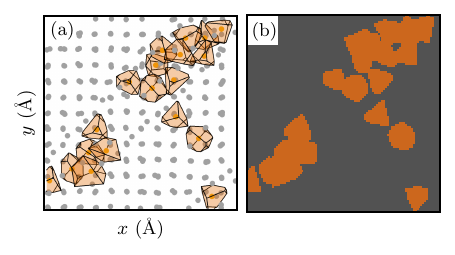}
	\caption{(a) A three dimensional Voronoi diagram showing the oxygen volume in a Josephson junction simulated using molecular dynamics, showing the $xy$ plane which is perpendicular to the transport direction. Oxygen and aluminium atoms are indicated using orange and grey points respectively. The $z$ (transport) direction is into the page. (b) Results of surface coverage calculation for the same structure.}
	\label{fig:voronoi}
\end{figure}

In order to determine the percentage of substrate that is covered in oxygen for each of our structures we first determine the volume occupied by oxygen atoms in the barrier region. This is done by calculating the Voronoi cell around each atom in the barrier \cite{Voronoi1908, Anikeenko2004, [{ }][{, \texttt{http://control.ee.ethz.ch/$\mathtt{\sim}$mpt}}] MPT3}. Fig.~\ref{fig:voronoi} shows oxygen volumes for a low percentage coverage ($<$50\%) barrier on the `(100)1' crystal substrate.

For clarity, Fig.~\ref{fig:voronoi}~(a) shows the Voronoi cells for just the oxygen atoms in this structure, looking down on the substrate. By projecting the 3D Voronoi cells onto a 2D plane we can determine a ratio of uncovered to covered regions, giving the percentage of the underlying substrate that is covered. Fig. \ref{fig:voronoi}~(b) shows a percentage coverage calculation indicating areas which are covered by an oxygen Voronoi cell and Table~\ref{tab:thickness} shows the percentage coverage values for each structure.

Comparing the various structures in Table~\ref{tab:thickness}, we see that percentage coverage and oxygen extent increase monotonically with number of oxygen atoms. However the barrier FWHM (which is key for defining the transport characteristics) can vary greatly from run to run, and as a function of substrate crystal orientation. See for example, structure `(100)2' in the range 200-300 oxygen atoms. This is because the barrier characteristics depend strongly on the microstructure details of the oxide region. Note that we do not set the thickness or the percentage coverage of a simulated junction, rather these are emergent properties of the simulation.

\begin{figure}[htb]
	\centering
	\includegraphics[scale=1]{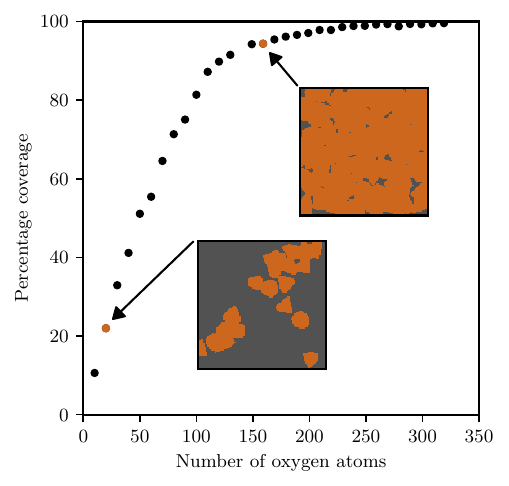}
	\caption{The oxidation process in a molecular dynamics simulation of AlO$_x$ growth on the `(100)1' aluminium substrate before overgrowth with a metal contact. Data points are shown for every tenth oxygen atom that is bound to the surface. Results show the percentage of the aluminium substrate that is covered by oxygen atoms as a function of the number of oxygen atoms in the insulating barrier with a cross-sectional area of 576 {\AA}$^2$.}
	\label{fig:oxide_formation}
\end{figure}

Fig.~\ref{fig:oxide_formation} shows the relationship between the substrate coverage against the number of oxygen atoms in the barrier for the `(100)1' structure. Slices are taken at every tenth oxygen atom that is bound to the surface during oxidation in the molecular dynamics simulation.

In Fig.~\ref{fig:oxide_formation} we see two different regimes. Below ${\sim}100$ oxygen atoms (per $24$ {\AA} $\times 24$ {\AA} area) we see one gradient which we attribute to individual oxygen atoms being deposited onto the metal substrate, and above ${\sim}100$ oxygen atoms we see a different gradient attributed to the formation of an oxide layer. The insets show the Voronoi surface coverages for two structures within each regime as calculated in Fig.~\ref{fig:voronoi}. These results indicate that below  ${\sim}100$ oxygen atoms is where pinhole formation is likely to occur before we see behaviour consistent with a film of AlO$_x$ as the number of oxygen atoms increases.

\section{Normal transport in atomistic model} \label{sec:md_transport}
Having developed an atomistic model of Al$-$AlO$_x-$Al tunnel junctions using molecular dynamics we now use this model to obtain information about the transport properties of these 3D simulated junctions. The transport calculations in this section do no include superconducting parameters or considerations of Cooper pairs of electrons. They describe the device in its normal state and are purely single-electron transport calculations.

\begin{figure}[H]
	\centering
	\includegraphics[scale=1]{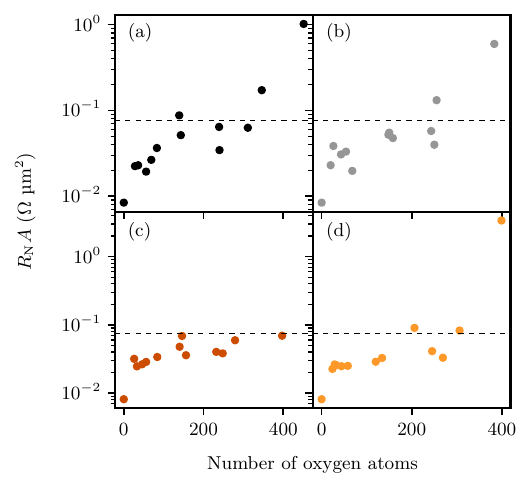}
	\caption{Resistance area for Al$-$AlO$_x-$Al junctions with increasing numbers of oxygen atoms in the barrier. Each panel shows data for structures grown on different aluminium substrates, and two calculations of each. The structures for each panel are (a) `(100)1', (b) `(100)2', (c) `(111)1', (d) `(111)2'.}
	\label{fig:pinhole_resistance}
\end{figure}

We use a NEGF method to calculate the transmission and the normal resistance, $R_\textrm{N}$, in the same way as we did for the toy model in Sec.~\ref{sec:toy_model}. Fig. \ref{fig:pinhole_resistance} shows the resultant $R_\textrm{N}A$ relationship for junctions grown on each substrate. As we expect $R_\textrm{N}A$ to be exponential with film thickness \cite{Dorneles2003}, results are plotted on a logarithmic scale.

On each of the data sets in Fig.~\ref{fig:pinhole_resistance} the dashed line indicates the resistance quantum that corresponds to $T = 1$ as determined in Fig.~\ref{fig:toy_transport}. It corresponds to the expected transition from metallic to insulating behaviour. For every data set we see significant variation in the resistance values, and no clear transition to insulating behaviour until ~400 oxygen atoms are deposited. An exception to this is in the `(111)1' data set where we find that the last data point is below the metal-to-insulator transition even with approximately 400 oxygen atoms in the barrier, like the other structures. This implies there may be more metallic regions present in the barrier in the form of pinholes or weak-points.

In this study we aim to create a distinction between pinholes and weak-points. We define pinholes to be present when the Al substrate is $<$100\% covered in oxide, which results in a direct metallic link once metal is deposited as a top contact (as in Sec. \ref{sec:toy_model}). We define weak-points to be present when our surface coverage calculations show no presence of pinholes, however the transport calculations indicate a localised higher current density compared to a uniform barrier. Such localised variation in the current density results from variation in the oxide density or stoichiometry but may not correspond to the conventional image of a pinhole, as illustrated in Fig.~\ref{fig:pinhole}.

Comparing these results to Fig.~\ref{fig:oxide_formation}, we find that an increase in surface coverage (especially close to 100\%) isn't directly correlated to insulating behaviour in the barrier. We reach close to 100\% coverage when approximately 200 oxygen atoms are deposited, however for all data sets in Fig.~\ref{fig:pinhole_resistance} we see that the $R_\textrm{N} A$ data points around 200 oxygen atoms are close to or below the metal-to-insulator transition, indicating metallic behaviour. This highlights that surface coverage alone cannot give us a good indication of a functioning barrier.

The variation we see in the resistance data in Fig.~\ref{fig:pinhole_resistance} is consistent with measurements in manufactured Al$-$AlO$_x-$Al tunnel junctions \cite{Kreikebaum2020,Osman2021,Verjauw2022}. For example, it has been shown that when some device barriers are grown to thicknesses which are typically expected to display insulating behaviour, a few may fail due to an excess current \cite{Tolpygo2008}. Dataset `(111)1' shows a theoretical example which could explain these situations. As well as this, experiments have shown significant variability in the transport between devices fabricated in the same way. We believe these variabilities may arise from the disordered nature of the AlO$_x$ region as results in Fig.~\ref{fig:toy_transport} showed significantly less variability, likely due to the uniform potential barrier used and systematic changes to the pinhole size. It has also previously been shown that crystal grain boundaries in the aluminium bottom contact could cause thickness variations in the AlO$_x$ layer that is grown above \cite{Fritz2019}. These thickness variations could then in turn cause irregularities in the transport measurements. In this study we have not included the effects of grain boundaries, we grow devices on single crystal aluminium substrates. Despite omitting grain boundaries, we still see significant variability, highlighting that disorder alone may be a key contributor to the variation in transport we see in junctions of this sort.

There also appears to be a larger spread in the data for junctions grown on Al (100) substrates compared to Al (111). This could indicate that the spatial variation of the junction resistance for the (111) crystal orientation is lower, however further statistics are required to be definitive. The computational expense of growing large atomic structures and performing atomic-resolution transport calculations limits our ability to perform detailed replication studies at this time.

In prior work by Cyster \etal \cite{Cyster2020} a reliable exponential trend was observed in $R_\textrm{N}A$ calculations for junctions with increasing thickness. That work used a melt and quench method to create devices which gave control over the oxide thickness, however it did not consider the experimental fabrication processes used. In this study we mimic the fabrication process more closely \cite{Cyster2021} aiming to provide a more accurate representation of what would be obtained from measurements of real devices. However, this does produce greater variability in the transport results. Cyster \etal \cite{Cyster2020} also uses a S-M potential, whereas this work uses a ReaxFF potential which has since been shown to more accurately describe the self-limiting behaviour we see experimentally \cite{Cyster2021}.

\begin{figure}[b]
	\centering
	\includegraphics[scale=1]{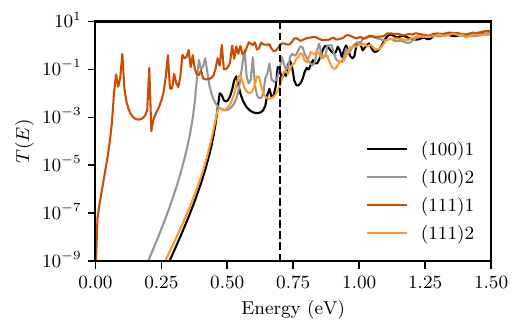}
	\caption{Transmission functions for the thickest barriers in each of our data sets. The dashed line indicates $E_\textrm{F} = 0.7$ eV which is used for the calculations in this work.}
	\label{fig:pinhole_transmission}
\end{figure}

To investigate the variation we see in the transport calculations, Fig.~\ref{fig:pinhole_transmission} shows the transmission function for each of the thickest structures, corresponding to more than 350 oxygen atoms in the simulation cell.

In particular, Fig.~\ref{fig:pinhole_transmission} shows the sensitivity of transport measurements to changes in the Fermi energy, noting that our calculations all use $E_\textrm{F} = 0.7$ eV, marked on Fig.~\ref{fig:pinhole_transmission} with a dashed line. From the transmission we also see that the `(111)1' structure is more transparent than the other three structures, supporting results from Fig.~\ref{fig:pinhole_resistance} where we found that the thickest structure was metallic compared to the junctions in the other data sets.

Our calculations discussed up to this point give us an indication of the overall transport in the junction. However to understand the specific effects that pinholes have on transport within the barrier we need to study how current flows through the junction spatially. To investigate this we calculate the three-dimensional current density throughout the junction.

The current flowing between two points in a junction is given by the element-wise product of $H$ and $G^n$,
\begin{equation}
J(\mathbf{r}, \mathbf{r'}, E) = \frac{e}{h} \textrm{Im}[H \circ G^n]
\end{equation}
where the electron Green's function is given by
\begin{equation}
G^n(E) = G^r \Sigma^{in} G^a \:,
\end{equation}
and
\begin{equation}
\Sigma^{\textrm{in}}(E) = \Gamma_\textrm{S} f_\textrm{S} + \Gamma_\textrm{D} f_\textrm{D} \:.
\end{equation}

Using  $G^n(E)$, the charge density can be calculated in three-dimensions with
\begin{equation}
n(x,y,z) = \frac{1}{2 \pi a_x a_y a_z} \textrm{diag}(G^n(E)) \:.
\end{equation}

The current in one direction is calculated from the difference between pairs of points, for the $z$ direction this looks like
\begin{equation}
J_z(x,y,z;E) = \frac{1}{a_x a_y}[J(\mathbf{r}, \mathbf{r'}, E) - J(\mathbf{r'}, \mathbf{r}, E)] \:.
\end{equation}

\begin{figure}[H]
	\centering
	\includegraphics[scale=1]{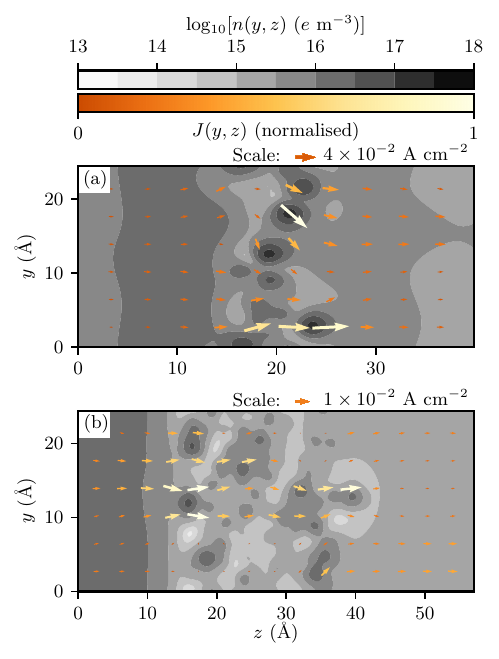}
	\caption{Orange arrows show the current density through the `(111)1' device for (a) the lowest coverage structure with \#~O~=~26 (pinholes), and (b) the thickest barrier (comparable to a fabricated device) with \#~O~=~397. Charge density throughout the junction is shown in greyscale. Orange arrows show the localised current densities. The arrow scale indicates the magnitude of the current density and the arrow colour table is normalised.}
	\label{fig:current_pinhole}
\end{figure}

Fig. \ref{fig:current_pinhole} shows $yz$ slices of the barrier and the charge and current densities for positions in those slices.

Dark regions in Fig.~\ref{fig:current_pinhole} correspond to areas in the junction which have a higher charge density, typically indicative of a metallic region. The orange arrows show the single-electron current density at positions in the junction. In both figures it is apparent that the largest proportion of current flows through localised regions of the junction. Fig.~\ref{fig:current_pinhole}~(a) shows the lowest surface coverage junction in the `(111)1' data set (26 oxygen atoms), and Fig.~\ref{fig:current_pinhole}~(b) shows the same substrate but for the thickest oxide for full surface coverage (397 oxygen atoms). The current in Fig.~\ref{fig:current_pinhole}~(a) is on average approximately four times higher than in Fig.~\ref{fig:current_pinhole}~(b) as is expected for a more transparent barrier. Of particular interest, despite the device in Fig.~\ref{fig:current_pinhole}~(b) having a thick enough barrier to expect insulating behaviour, there are still weak-points which facilitate peaks in the single-electron current flow. 

The `(111)1' structure shown in Fig.~\ref{fig:current_pinhole}~(b) is the same device that has displayed a higher transparency in Fig.~\ref{fig:pinhole_transmission} and Fig.~\ref{fig:pinhole_resistance}~(c). Fig.~\ref{fig:current_pinhole} allows us to probe how the microscopic structure of the device may contribute to the higher transparency across the junction, emphasising that the disordered barrier may be a crucial consideration when trying to predict the parameters of fabricated Josephson junction devices. 

In general, defects in a superconducting qubit can couple via charge, flux, or critical current in the circuit \cite{Cole2010}.  In the case that defects in the junction modulate the critical current \cite{Constantin2007, Muller2019}, the qubit-defect coupling strength is proportional to the magnitude of the (normal or super) current flow at the site of the defect. Therefore, increased variance in the current density through the junction can result in increased coupling to defects.

\begin{figure}[b]
	\centering
	\includegraphics[scale=1]{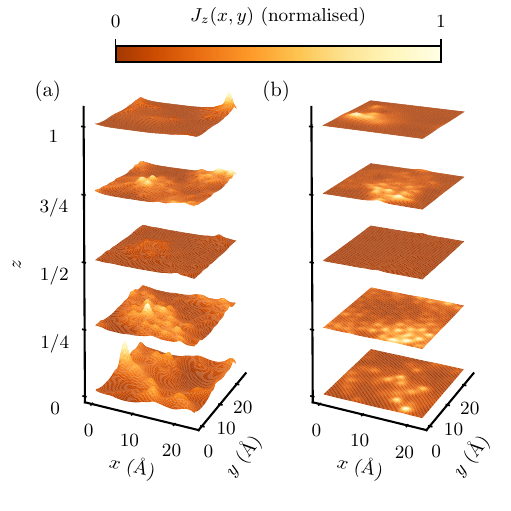}
	\caption{Current density $xy$ slices through the transport direction, $z$, for the thickest barriers in (a) the `(100)1' and (b) `(111)2' device. Slices are taken at regular $z$ intervals through the barrier region as fractions of the oxygen extent in each junction. Current densities are normalised by the maximum in the `(100)1' data set.}
	\label{fig:current_density_spot}
\end{figure}

Fig.~\ref{fig:current_density_spot} shows $xy$ slices of the magnitude of the current density perpendicular to the transport direction for the thickest barriers in `(100)1' and `(111)2' data sets. For figures that follow we compare the `(100)1' and `(111)2' data sets where the thickest two barriers are both insulating.

The slices shown in Fig.~\ref{fig:current_density_spot} are taken at equal intervals through the oxygen extent of the barrier region. The distribution of the current density between slices varies greatly, specifically we see that the current peaks in different positions between slices. This highlights that for thicker structures without clear pinholes in the oxide, we find weaker, indirect paths of single-electron current (weak-points). We suspect this is due to aluminium rich regions in the oxide enabling a current flow with an indirect path. It is important to note that the edges of the oxide are not abrupt as there is significant oxygen diffusion into the contacts \cite{Cyster2021, Zeng2016}. This stoichiometric gradient results in more current hotspots at the starts and ends of the oxide compared to the middle of a junction. The central slices from Fig.~\ref{fig:current_density_spot} are then the most indicative of the behaviour of the insulating barrier in the junction.

Although these weak-points have the most significant effect in `(111)1' data set, they are nonetheless present in all junctions studied in this work.

\section{Superconducting transport in atomistic model}
\label{sec:superconducting}
Higher transparency channels within the tunnel barrier in Josephson junctions can cause variations in the transmission dependent CPR \cite{Willsch2024}. The results in Sec.~\ref{sec:md_transport} indicate higher transparency regions in the normal current density. Following the same methods used in Sec.~\ref{sec:toy_sc} for the toy model, we determine the transmission eigenvalues and CPR for each structure.

In this section we also follow recent work which shows that the contributions from different transmission channels can be written as a Fourier series \cite{Willsch2024}:
\begin{equation}
I_\textrm{S}(\phi) = \sum_{m=1}^{\infty} c_m \sin(m\phi) \:,
\label{eq:fourier_cpr}
\end{equation}
where $c_m$ are the transmission dependent Fourier coefficients of order $m$.

The ratio of the fundamental sinusoidal signal to the higher order contributions is given by
\begin{equation}
\nu = \frac{1}{c_1} \sum\limits_{m=2}^{\infty} c_m \:,
\end{equation}

which allows us to quantify how sinusoidal the CPR is.

\begin{figure}[H]
	\centering
	\includegraphics[scale=1]{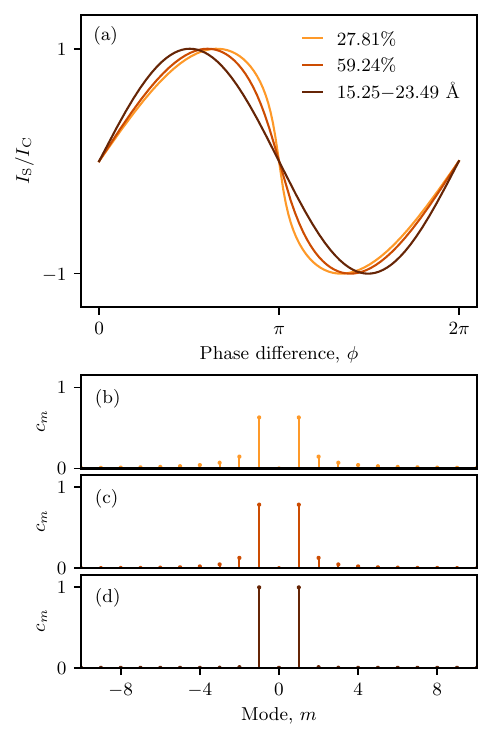}
	\caption{(a) CPRs for structures within the `(100)1' data set, as labelled in the legend. Each curve is normalised by the critical current, $I_\textrm{C} = $0.169 \textmu A, 0.132 \textmu A, 0.003 \textmu A for the 27.81\%, 59.24\% and 15.25-23.49 {\AA} structures respectively, all of which have a cross-sectional area of 24$\times$24~{\AA}$^2$. (b,c,d) The frequency domain representation of each of the curves in (a), highlighting the different sinusoidal contributions to each signal. The Fourier coefficients, $c_m$, are normalised such that their sum equals one.}
	\label{fig:fourier_cpr}
\end{figure}

Fig.~\ref{fig:fourier_cpr} shows the CPR and Fourier components calculated for three structures within the `(100)1' data set.

\begin{figure}[b]
	\centering
	\includegraphics[scale=1]{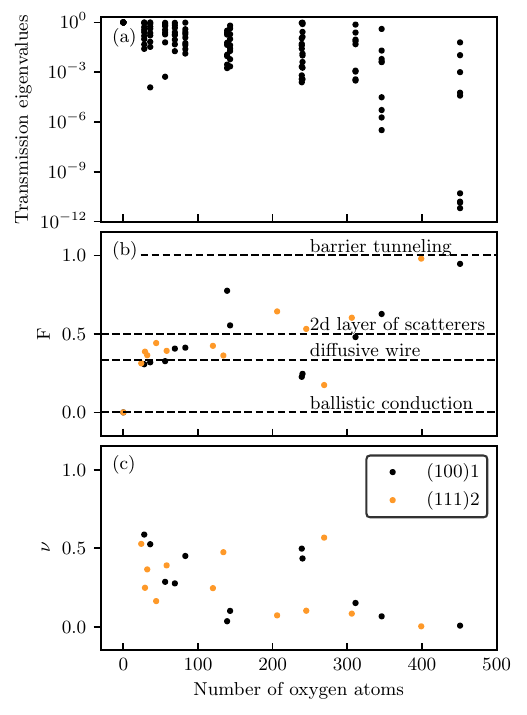}
	\caption{(a) Eigenvalues of the transmission matrix for the `(100)1' data set. (b) Fano factors for each structure in the `(100)1' data set (black) and `(111)2' data set (orange), dashed lines indicate different types of nanoscale conduction from the literature. (c) The ratio of higher order Fourier coefficients to the fundamental coefficient. A value of zero corresponds to a purely sinusoidal signal.}
	\label{fig:eigs_fano}
\end{figure}

The magnitudes of the CPRs in Fig.~\ref{fig:fourier_cpr}~(a) differ by orders of magnitude. For this reason we choose to scale them by the critical current, $I_\textrm{C}$, which is equivalent to the maximum for each curve. These values are given in the caption and vary between structure. We find the structures with the least amount of oxygen to have a higher $I_\textrm{C}$ compared to the thickest barriers. Scaling the results in Fig.~\ref{fig:fourier_cpr}~(a) allows us to more clearly study the shape of the CPR. For lower surface coverages we see a deviation from the expected sinusoidal Josephson relationship, $I_\textrm{S}(\phi) = \sin \phi$ \cite{Willsch2024, Golubov2004}. When we look at the Fourier transform of each $I_\textrm{S}$ curve we find that for junctions with 27.81\% and 59.24\% coverage there are extra contributions in the Fourier series expansion in Eq.~\ref{eq:fourier_cpr}. These arise from the higher transmission probability eigenvalues for the thinner barrier structures, and result in a deviation from the sinusoidal relationship. To show this clearly, Fig.~\ref{fig:fourier_cpr} (b), (c), and (d) give a breakdown of different sinusoidal contributions to the CPR. Most notably, the thickest junction shown in Fig.~\ref{fig:fourier_cpr}~(d) has only one significant contribution, the main sinusoidal signal, and is the closest to an ideal Josephson junction. As discussed in Sec.~\ref{sec:toy_sc}, although we see a systematic change toward sinusoidal behaviour in Fig.~\ref{fig:fourier_cpr}, this is not necessarily indicative of every structure in the data set. The shape of the CPR is dependent on the number of channels with transmission probabilities close to 1, which in turn depends on the structure of the oxide and the presence of pinholes. Fig.~\ref{fig:eigs_fano} summarises data from Fig.~\ref{fig:fourier_cpr} for all the structures in the `(100)1' and `(111)2' data sets, including the transmission eigenvalues for the `(100)1' structure. 

Fig.~\ref{fig:eigs_fano}~(a) shows the `(100)1' set of data for the eigenvalues of the transmission matrix calculated from the Green's function. We see that for thinner barriers the eigenvalues are closer to 1, indicating a more transparent junction. As the barriers become thicker the eigenvalues begin to drop to zero. The Fano factor, shown in Fig.~\ref{fig:eigs_fano}~(b) is given by
\begin{equation}
F = \frac{\sum_{n} T_n (1-T_n)}{\sum_{n} T_n} \:,
\end{equation}

which allows us to differentiate between different types of nanoscale conduction \cite{Blanter2000}. We find that as we grow thicker barriers the Fano factor approaches 1, indicating barrier tunnelling, and for a pure aluminium we obtain a Fano factor of 0 corresponding to ballistic transport as indicated on the graph with dashed lines. A Fano factor of 1/3 corresponds to a diffusive wire, and 1/2 a 2D layer of randomly distributed scatterers \cite{Heikkila2013, Blanter2000}. We note that many of the points in the thinner barriers are situated around the diffusive wire regime, and we do also see a few points near the 2D layer of scatterers regime. However, given the variability in our data we cannot distinguish between these two regimes definitively.

Due to the logarithmic scale in Fig.~\ref{fig:eigs_fano}~(a) for the `(100)1' data set, the results in Figs.~\ref{fig:eigs_fano}~(b) and (c) for the same data set may show a greater discrepancy (by eye) compared to the eigenvalues in Fig.~\ref{fig:eigs_fano}~(a). Fig.~\ref{fig:eigs_fano}~(c) shows the ratio of the largest sinusoidal contribution to the CPR with the other contributions, highlighting results in Fig.~\ref{fig:fourier_cpr} which indicate that thicker barriers approach a pure sinusoidal CPR and a $\nu \simeq 0$, indicative of an ideal Josephson junction. For the `(100)1' data set, the junction with the strongest sinusoidal response in Fig.~\ref{fig:eigs_fano}, with $\nu = 0.0071$, has a thickness between 15.25~{\AA} and 23.49~{\AA} based on the FWHM and O extent in Table~\ref{tab:thickness}. Fig.~\ref{fig:eigs_fano} indicates that lower oxygen content in the barrier region leads to channels with higher transmission probabilities, resulting in a deviation from the expected sinusoidal response. This is particularly true for thinner barriers where pinholes are more likely to be present. In Sec.~\ref{sec:md_transport} we found an anomaly in the thickest barrier for the `(111)1' data set, where there were no pinholes present however there were weak-points in the barrier. The term weak-points is synonymous with channels that have higher transmission probabilities.

\begin{figure}[H]
	\centering
	\includegraphics[scale=1]{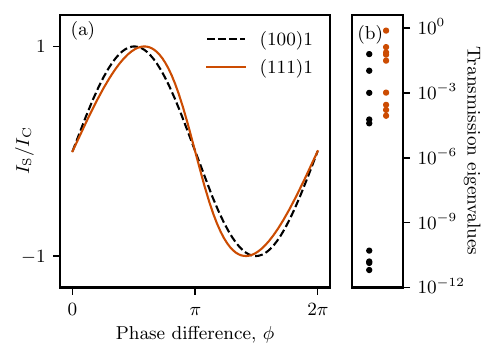}
	\caption{(a) CPRs for the thickest structures in the `(100)1' and `(111)1' data sets, each curve is normalised by the critical current. (b) The transmission matrix eigenvalues for the same two structures. For the `(100)1' and `(111)1' structure respectively, $\nu = 0.0071$ and $\nu = 0.1777$.}
	\label{fig:thickest_cpr}
\end{figure}

Fig.~\ref{fig:thickest_cpr} directly compares the `(111)1' data set to the `(100)1' for the thickest barriers. Fig.~\ref{fig:thickest_cpr}~(b) shows the larger proportion of higher transmission eigenvalues for the `(111)1' data set compared to the `(100)1' data set. As we found for junctions with pinholes, these higher transmission channels result in extra contributions to the CPR given in Eq.~\ref{eq:fourier_cpr}. Fig.~\ref{fig:thickest_cpr}~(a) shows how this results in a deviation from the expected sinusoidal response, which `(100)1' shows an example of based on its $\nu$ ratio given in Fig.~\ref{fig:eigs_fano}~(c) and in the caption of Fig.~\ref{fig:thickest_cpr}. How pure the sinusoidal response is has a direct influence on the effective Hamiltonian in superconducting qubits. To accurately calibrate qubits it is essential to be able to predict the CPR \cite{Willsch2024}. Our results indicate that both pinholes and weak-points in the barrier can lead to corrections to the conventional sinusoidal response.

\section{Conclusion}
The results of this work show the key role that both pinholes and disorder in the amorphous oxide barrier play in Al$-$AlO$_x-$Al tunnel junctions. Prior work has shown variations in the critical current of Josephson junction devices fabricated in the same way. Using molecular dynamics we have been able to probe the microscopic structure of these devices and analyse the oxidation process step-by-step. We have shown that in thinner barriers pinholes may be present and be a source of single-electron current which could in turn affect the transport properties of Josephson junction devices. Further to this we have also observed a significant amount of variability in junctions with a complete barrier which appears to be caused by the inherently disordered oxide barrier. To understand the minimum amount of oxide required to have a functioning device we studied the metal-insulator transition and found for a uniform electric potential we should see a transition to insulating behaviour between 90-100\% coverage. In reality the amorphous structure of the oxide creates a non-uniform electric potential which makes this transition less clear. In fact even when barriers are grown to have thicknesses comparable to experimentally fabricated devices, we might not see a transition to insulating behaviour which could be a contributing factor to device failure. These devices do not have pinholes in the barrier, but have weak-points which can facilitate single-electron current flow.

We find that the variations in transport that arise from the non-uniform electric potential also have consequences on the CPR. More transmissible channels due to pinholes result in more contributions to the CPR, causing it to deviate from the expected pure sinusoidal signal. As barriers are grown thicker we obtain a stronger sinusoidal signal. Although this is the case, we also find that weak-points in the barrier (in addition to fully metallic pinholes) can lead to corrections to the conventional sinusoidal response. This emphasises it is not only essential that a Josephson junction be free of pinholes but also that the oxide barrier be as uniform as possible. Such precise control of junction thickness, density, and stoichiometry may ultimately only be possible with epitaxial junctions \cite{Muller2019,Kline2012,Patel2013}.

Previously, the specific processes of oxide growth and why thinner films may not function as expected were not well understood. Our work uses computational models to analyse the oxidation process and the formation of structures that may cause faults in junctions with oxide barriers. The analysis in this work aims to inform and support Al$-$AlO$_x-$Al device fabrication.

\section{Acknowledgements}
The authors acknowledge the support of the Australian Research Council through grants CE170100039, and the National Computation Infrastructure (NCI) which provided the computational resources to undertake this work. The authors also acknowledge useful discussions with B.~Muralidharan and I.~Pop, and assistance of L.~Moshovelis with the creation of Fig.~\ref{fig:pinhole}. We acknowledge the people of the Woiwurrung, Wathawurrung and Boonwurrung language groups as the traditional custodians of the land on which this research was conducted, and the people of the Ngunnawal language group who are the Traditional Custodians of the land of the Australian Capital Territory where NCI is located. 

\newpage
\bibliography{pinhole_bibliography}

\end{document}